\documentclass[12pt,epsfig]{article}

\usepackage{slashed}
\usepackage{graphicx}
\usepackage{amsmath}
\usepackage{placeins}

\newskip\humongous \humongous=0pt plus 1000pt minus 1000pt
  \newif\ifdtup

\def\frac#1#2{ {{#1} \over {#2} }}

\def\eg{\hbox{\em e.g. }}
\def\ie{\hbox{\em i.e.}}

\def\beq{\begin{equation}}
\def\eeq{\end{equation}}

\def\beqn{\begin{eqnarray}}
\def\eeqn{\end{eqnarray}}

\def\MSbar{\overline{\rm MS}}

\def\LL{{l}}

\begin{document}

\title
{                   
High-loop perturbative renormalization constants \\
for Lattice QCD (III): \\
three-loop quark currents for Iwasaki 
gauge action and $n_f=4$ Wilson fermions}

\author
{M.~Brambilla, F.~Di~Renzo and M.~Hasegawa$^\dagger$ \\
\small{Dipartimento di Fisica e Scienze della Terra, Universit\`a di Parma} \\
\small{and INFN, Gruppo Collegato di Parma} \\
\small{I-43100 Parma, Italy} \\
$^\dagger$ \small{{\em current address} Bogoliubov Laboratory of
  Theoretical Physics, Dubna, Russia}
}

\maketitle

\begin{abstract}
This is the third of a series of papers on three-loop
computation of renormalization constants for Lattice QCD. 
Our main point of interest are  
results for the regularization defined by Iwasaki gauge 
action and $n_f=4$ Wilson fermions. Our results for quark bilinears
renormalized according to the RI'-MOM scheme 
can be compared to non-perturbative results. The latter are available
for Twisted Mass QCD: being defined in the chiral limit,
renormalization constants must be the same. \\
We also address more general problems. In particular, 
we discuss a few methodological issues connected to summing the
perturbative series such as the effectiveness of Boosted Perturbation
Theory and the disentanglement of irrelevant and finite
volume contributions. Discussing these issues we consider not only the
new results of this paper, but also those for the regularization 
defined by tree-level
Symanzik improved gauge action and $n_f=2$ Wilson fermions,
which we presented in a recent paper of ours. 
We finally comment to which extent the techniques we
put at work in the NSPT context can provide a fresher look into the
lattice version of the RI'-MOM scheme.  
\end{abstract}

\section{Introduction}

Numerical Stochastic Perturbation Theory (NSPT \cite{NSPT0,NSPT1}) can
be a powerful tool to address perturbative computations in Lattice QCD
up to an order which would be impossible to attain with standard,
diagrammatic approaches. A few years ago the Parma group applied NSPT to get
three- (and even four-) loop Renormalization Constants (RCs) of finite quark bilinears 
in the scheme defined by Wilson gauge action and Wilson fermions 
\cite{NSPT_Zs}. Very recently, \cite{NSPT_TLS} provided in turn both 
finite and logarithmically divergent three-loop
RCs for currents in the scheme defined by 
tree-level Symanzik improved gauge action and two flavors of 
Wilson quarks. The inclusion of divergent RCs was made possible by the
method first introduced in \cite{NSPT_Gh,NSPT_Gl}: when an anomalous
dimension is in place, finite size effects can be important in NSPT
computations and they have to be carefully taken into account. 
The main result of the current paper is the computation of 
quark currents RCs in the regularization defined by Iwasaki
gauge action and four flavor of Wilson quarks 
(quenched computations will be reported as well, to enable a comparison). 
Preliminary results were quoted in \cite{NSPT_Zs_LAT2012}. 
For a complete discussion of our methodology the
reader should refer to \cite{NSPT_TLS}, which has 
been largely devoted to discuss in some detail the NSPT approach to
the computation of renormalization constants (with a main emphasis on
the control over finite lattice spacing and finite volume effects). 

Both in the case of $n_f=2$ tree-level Symanzik
gauge action and in the case of $n_f=4$ Iwasaki gauge action, our results 
can be compared with analogous non-perturbative computations for 
Twisted Mass fermions 
\cite{ETMC_Zs, ETMC_ZsIWA} (the renormalization scheme is 
massless and thus RCs are the same). 
In order to do that, perturbative series have to be summed. An 
important goal of this paper is a 
discussion of the issues that are related to summing PT series for
Lattice QCD. 

\noindent The overall structure of this paper is as follows:
\begin{itemize}
\item Section \ref{sec:3loopZ} presents an overview of our methodology. It
  is mainly intended to allow the reader to go through the paper
  without having to refer to other sources. 
\item In Section \ref{sec:res} our results for $Z_S$, $Z_P$, $Z_V$,
  $Z_A$ for Iwasaki gauge action and $n_f=0,4$ Wilson fermions are
  presented.
\item In Section \ref{sec:sum0} we address the issue of summing the
  series, and in particular we deal with the explicit disentanglement 
of irrelevant (finite $a$) contributions, which is possible once also
finite volume effects have been corrected for. We take into account
not only the results of the current paper, but also those of 
\cite{NSPT_TLS} (\ie, we compare the two regularizations).
\item Section \ref{sec:sum} contains a discussion of different ways of
  summing the series. The effectiveness of Boosted Perturbation
Theory is discussed; it turns out that this is relevant in particular
for Symanzik action.
\item In Section \ref{sec:whatNEW} we briefly discuss to which extent our approach
  can provide a contribution for an overall better understanding of
  the lattice version of the RI'-MOM scheme.  
\end{itemize}

\section{Three loop renormalization constants in NSPT}
\label{sec:3loopZ}
In this section we provide a brief account of our computational strategy. 
This is basically a summary of the discussion of \cite{NSPT_TLS}, to 
which the interested reader is referred for an in-depth description of 
our method. 

\subsection{3-loop RI'-MOM lattice computations}

The lattice is a suitable regulator for the RI'-MOM renormalization
scheme \cite{RI-MOMrm}. The definition of the latter for quark currents 
starts from the computation of Green functions 
on external quark states at fixed momentum $p$
\[
G_{\Gamma}(p) \, = \, \int dx \,\langle p | \; \overline{\psi}(x) \Gamma \psi(x) \; | p \rangle. 
\]

\noindent The $G_{\Gamma}(p)$ are then amputated to get vertex functions ($S(p)$ is the quark propagator)
\[
\Gamma_{\Gamma}(p) = S^{-1}(p) \, G_{\Gamma}(p) \, S^{-1}(p). 
\]

\noindent By projecting on tree-level structure
\[
 O_\Gamma(p) = Tr\left(\hat P_{O_\Gamma}\Gamma_\Gamma(p)\right) 
\]
 one gets the quantities $O_\Gamma$ which enter the definition of the
currents renormalization constants 
\begin{equation}\label{eq:masterZ}
Z_{O_\Gamma}(\mu,\alpha) Z_q^{-1}(\mu,\alpha)O_\Gamma(p)|_{p^2=\mu^2} = 1.
\end{equation}
By choosing different $\Gamma$ one obtains the 
different currents, {\em e.g.} the scalar (identity),
pseudoscalar ($\gamma_5$),  vector ($\gamma_{\mu}$), axial
($\gamma_5\gamma_{\mu}$). The master formula (Eq.~(\ref{eq:masterZ})) 
is defined in terms of the quark field renormalization constants
which in turn reads
\begin{equation}\label{eq:Zq}
Z_q(\mu,\alpha) = -i\frac{1}{12}\frac{Tr(\slashed{p}S^{-1}(p))}{p^2}|_{p^2=\mu^2}.
\end{equation}
We adhere to the standard recipe of getting a mass-independent scheme
by defining everything at zero quark mass. \\

A main point in our strategy is to get the (divergent) logarithmic
contribution to the renormalization constants from continuum
computations: NSPT is only in charge of reconstructing the finite
parts. The typical renormalization constant we want to compute 
(in the continuum limit) reads
\begin{equation}\label{eq:Zexpans}
Z (\mu,\alpha) = 1+\sum_{n>0}d_n(\LL)\,\alpha(\mu)^n\qquad d_n(\LL)=\sum_{i=0}^nd_n^{(i)}\LL^i \qquad \LL\equiv\log(\mu a)^2
\end{equation}
where the lattice cutoff ($a$) is in place and the expansion is in the renormalized coupling. 
Divergencies can show up as powers of $\LL = \log(\mu a)^2$. By
differentiating Eq.~(\ref{eq:Zexpans}) with respect to $\LL$ 
one obtains the expression for the anomalous dimension 
\[
\gamma = \frac{1}{2} \frac{d}{d\LL} \log Z
\] 
whose expansion can be read 
from continuum computations \cite{Gracey} 
\begin{equation}\label{eq:AnomDim}
\gamma = \sum_{n>0}\gamma_n \,\, \alpha(\mu)^n.
\end{equation}
This is a scheme dependent, finite quantity, with no 
dependence on the regulator left. By imposing that the expression we
get by differentiating Eq.~(\ref{eq:Zexpans})
matches Eq.~(\ref{eq:AnomDim}) 
we can obtain the expressions of all the $d_n^{(i >0)} \, (n\leq 3)$, 
which are thus expressed in terms
of the $\gamma_{m\leq n}$, the $d_{m \leq n}^{(0)}$ and the 
coefficients of the $\beta$-function; the latter come into place 
since part of the dependence on $\mu$ in Eq.~(4) is via the 
coupling $\alpha(\mu)$.\\

In the above discussion there was no reference to a (covariant) gauge
parameter $\lambda$. This is legitimate, since we compute in Landau
gauge, \ie $\;\lambda=0$. In a generic (covariant) gauge, one has a 
dependence on $\lambda$ entering Eq.~(\ref{eq:Zexpans}). Moreover, 
the gauge parameter anomalous
dimension comes into place in linking Eq.~(\ref{eq:Zexpans}) to 
Eq.~(\ref{eq:AnomDim}). Since the non trivial dependence
on the gauge parameter anomalous dimension is itself proportional to
$\lambda$, all this is immaterial in Landau gauge: if one keeps
track of all the $\lambda$-dependence and then puts 
$\lambda=0$ one gets the same result which is got by ignoring 
$\lambda$ from the very beginning. \\

In our computations the Zs are expressed as expansion in the bare
lattice coupling $\alpha_0$
\begin{equation}\label{eq:Z_al0}
Z (\mu,\alpha_0) = 1+\sum_{n>0}\overline{d}_n(\LL)\,\alpha_0^n \qquad \overline{d}_n(\LL)=\sum_{i=0}^n\overline{d}_n^{(i)}\LL^i.
\end{equation}
Eq.~(\ref{eq:Z_al0}) is obtained from Eq.~(\ref{eq:Zexpans}) by plugging
into the latter the matching of the renormalized coupling to the
lattice bare one.

\subsection{2-loop matching of $\alpha_{\mbox{\tiny{IWA}}}$ to continuum}

The matching of $\alpha_{\mbox{\tiny{IWA}}}$ to a continuum coupling
is only known to one-loop \cite{Aoki}. Since we need a two-loop
matching to get Eq.~(\ref{eq:Z_al0}), we had to compute it. This
was done by first matching $\alpha_{\mbox{\tiny{IWA}}}$ to an
intermediate scheme, which was chosen to be a potential scheme. 
The matching of the latter to $\MSbar$ is known \cite{York} and the results for  
the anomalous dimension we can read from \cite{Gracey} are obtained as
expansions in $\alpha_{\mbox{\tiny{$\overline{MS}$}}}$. Thus, computing the matching
of potential coupling $\alpha_{\mbox{\tiny{V}}}$ to $\alpha_{\mbox{\tiny{IWA}}}$ is
a possible solution. Here and in
the following our notation only enlightens the dependence of the scheme
on the gluonic action: the dependence on Wilson fermions
has to be assumed as well, when we refer to the four flavors case. 
The strategy of the computation is that of \cite{residual_massQ,
  residual_massU}. The interested reader can find more information on
technical details both in \cite{NSPT_TLS} and in \cite{alphaMATCHlat2012}.

We started from the NSPT computation of Wilson loops $W(R,T)$ from which
we got Creutz ratios
\[
V_T(R) = \log\left(\frac{W(R,T-1)}{W(R,T)}\right).
\]
A potential for static sources at distance $r=Ra$ can now be defined
and a coupling out of it according to
\begin{eqnarray}\label{eq:ValphaLat}
a V(r) = a V(R a) & = & \lim_{T\to\infty}V_T(R) \\ \nonumber
& = & 2\delta m-C_F\frac{\alpha_V(r^{-1})}{R}
\end{eqnarray}
where one can see that in a lattice regularization a residual mass
$\delta m$ comes on top of the coupling. Here we need to rely on an
approximation, since we can not compute the limit in 
Eq.~(\ref{eq:ValphaLat}). As a consequence of the same observation,
our results are not in the continuum limit. Despite this, we could 
obtain a decent estimate of the matching that in perturbation theory 
reads ($r=Ra$)
\begin{equation}\label{eq:matchingAVA0}
\alpha_{\mbox{\tiny{V}}}(r^{-1}) = \alpha_{\mbox{\tiny{IWA}}} + C_1(R)\, \alpha_{\mbox{\tiny{IWA}}}^2+ C_2(R)\, \alpha_{\mbox{\tiny{IWA}}}^3 +
\mathcal{O}(\alpha_{\mbox{\tiny{IWA}}}^4)
\end{equation}
where the expansion coefficients are a function of scale parameters
$\Lambda$ and coefficients of the $\beta$-functions $b_i$
\begin{eqnarray}
\nonumber C_1(R) & = & 2b_0
\log\frac{\Lambda_{\mbox{\tiny{V}}}}{\Lambda_{\mbox{\tiny{IWA}}}}+2b_0\log{R}\label{eq:c1V0}\\ 
C_2(R) &=& C_1(R)^2+2b_1\log{R}+2b_1
\log\frac{\Lambda_{\mbox{\tiny{V}}}}{\Lambda_{\mbox{\tiny{IWA}}}}+\frac{b_2^{(\mbox{\tiny{V}})}-b_2^{(\mbox{\tiny{IWA}})}}{b_0}.\label{eq:c2V0}
\end{eqnarray}
Reconstructing one-loop result was a check that the procedure is
viable, and at two-loop we could finally obtain 
\begin{eqnarray}
\label{eq:ourX}
  \frac{b_2^{(\mbox{\tiny{V}})}-b_2^{(\mbox{\tiny{IWA}})}}{b_0} \equiv
  X & = & 13.7 \pm 1.6 \;\; (n_f=0) \nonumber \\
& = & 11.3 \pm 1.6 \;\; (n_f=4)
\end{eqnarray}
where the new piece of information is contained in the quantity $X$: 
in the following we will refer to the latter. 

\subsection{3-loop critical mass}

Staying at zero quark mass in our three-loop NSPT computation requires the
knowledge of the Wilson fermion critical mass at two-loop, which is
known from the literature \cite{HarisMC}. 

From now on, we switch to $\beta^{-1}$ as the expansion parameter for
our results. Also, we introduce a hat notation to denote dimensionless 
quantities, {\em e.g.} $\hat p = pa$ (if needed, explicit factors of
$a$ will be later singled out). 

The critical mass is computed from the inverse quark propagator 
($\hat{m}_W(\hat{p}) = {\cal O}(\hat{p}^2)$ is the irrelevant mass
term generated at tree level)
\beq
a\Gamma_2(\hat p, \hat m_{cr}, \beta^{-1}) = aS(\hat p, \hat m_{cr}, \beta^{-1})^{-1} 
= i\hat{\slashed{p}}+\hat m_{W}(\hat p) - \hat\Sigma(\hat p, \hat m_{cr}, \beta^{-1}).
\eeq
More precisely, in the self-energy
$\hat{\Sigma}(\hat{p},\hat{m}_{cr},\beta^{-1})$ 
we single out the components along the (Dirac space)
identity, the one along the gamma matrices and the irrelevant one 
along the remaining elements of the Dirac basis
\beq\label{eq:sigma}
\hat\Sigma(\hat p, \hat m_{cr}, \beta^{-1}) = \hat\Sigma_c(\hat p,
\hat m_{cr}, \beta^{-1})+\hat\Sigma_{\gamma}(\hat p, \hat m_{cr},
\beta^{-1})+
\hat{\Sigma}_{\rm other} (\hat p, \hat m_{cr}, \beta^{-1}),
\eeq
The critical mass can be read from
$\hat{\Sigma}_c$\footnote{This is by the way the reason for the
  subscript $c$.} at zero momentum 
\begin{equation}\label{eq:Mc}
	\hat{\Sigma}(0,\hat{m}_{cr},\beta^{-1}) = \hat{\Sigma}_c(0,\hat{m}_{cr},\beta^{-1}) = \, \hat{m}_{cr}.
\end{equation}
The known one- and two-loop values of the critical mass were inserted
(as counterterms): this is enough to have massless quarks at the order
we are interested in (three-loop). The novel three-loop result for the
critical mass is not relevant for the computations at hand: it is
simply a byproduct.

Computations were performed on different lattice sizes: $32^4$, 
$24^4$, $20^4$, $16^4$, $12^4$. 
Left panel of Fig.~1 shows the three-loop computation of 
$\hat{\Sigma}_c$ at different values of momentum on a $32^4$
lattice, in the $n_f=4$ case. We are interested in the zero momentum value, which can be
got by fitting our observable as an expansion in 
hypercubic invariants, \eg
\begin{equation}
\label{eq:Hinv}
\sum_\nu \hat p_\nu^2 \;\;\;\;\;\;\;
\frac{\sum_\nu \hat p_\nu^4}{\sum_\nu \hat  p_\nu^2} \;\;\;\;\;\;\;
{(\sum_\nu \hat p_\nu^2)}^2 \;\;\;\;\;\;\;
\sum_\nu \hat p_\nu^4 \;\;\;\;\;\;\;
\frac{\sum_\nu \hat p_\nu^6}{\sum_\nu \hat p_\nu^2} \;\;\;\;\;\;\;
\ldots
\end{equation}
This is a general feature of all our computations. On each lattice
size we got a different value and an infinite volume result could then
be obtained by extrapolation. Right panel of Fig.~1 displays results
plotted as a function of $N^{-2}$, which is the power that best fits
our data. $N=L/a$ is the only significant quantity in the NSPT
context (there is no value one can attach to the lattice spacing
$a$). The infinite volume extrapolation was first fitted by keeping 
only the single power $-2$. 
We then checked that the central values of fits performed 
adding other powers were consistent with that result, within the error
of the latter. This procedure, thought limited by the number of
available sizes, proved to be accurate enough, as we could check at
one- and two-loop level, for which the expected zero value of the
critical mass was obtained in the infinite volume limit. Our final
results for the three-loop contribution to the critical mass are 
$\hat{m}_{cr}^{(3)} = -0.98(1) \;\; (n_f=0)$ and 
$\hat{m}_{cr}^{(3)} = -0.78(2) \;\; (n_f=4)$.

\begin{figure}[t]
\begin{center}
  \begin{tabular}{cc}
     \includegraphics[height=6.5cm,clip=true]{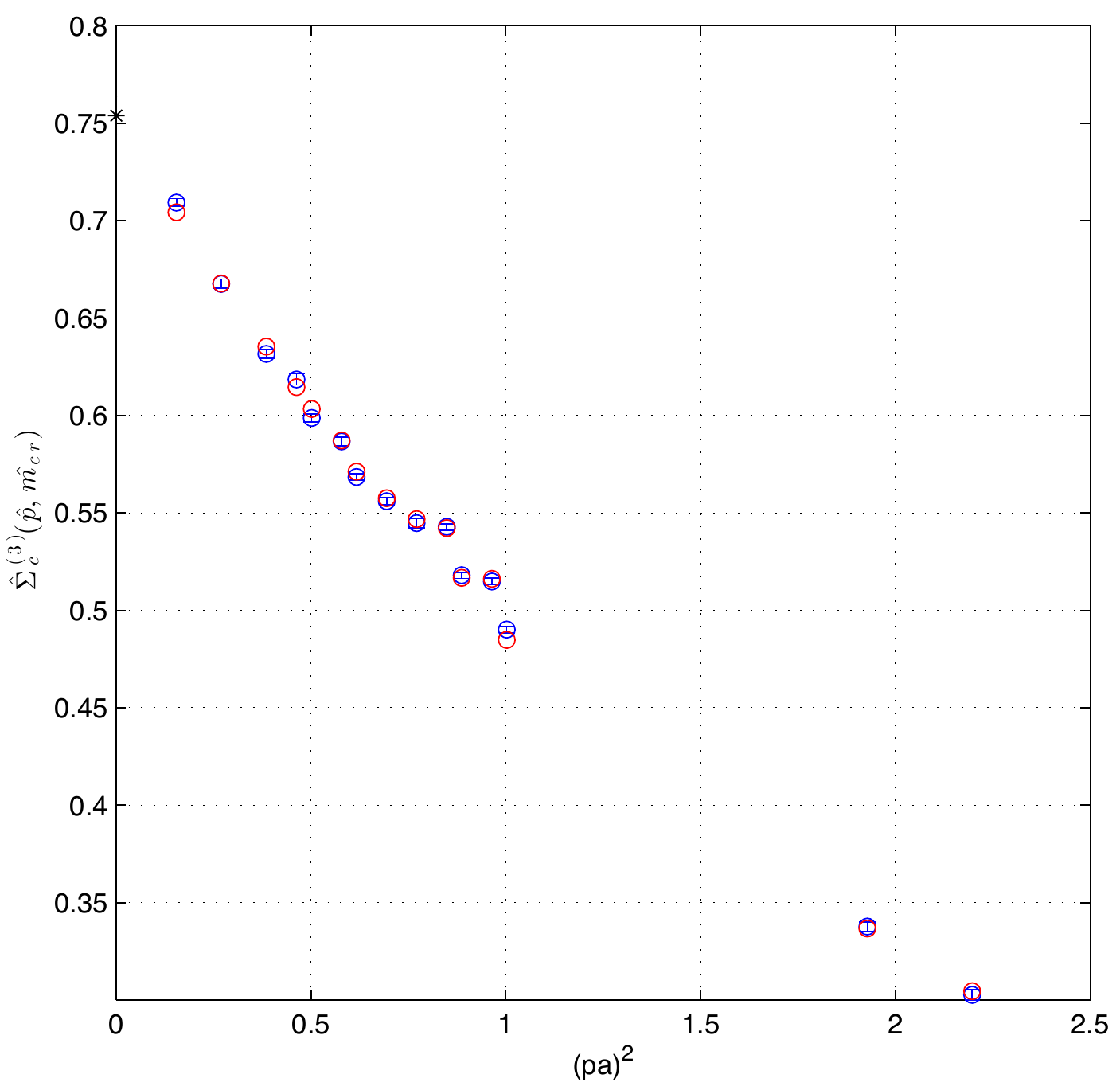}
     &
 \hspace{1.cm}
     \includegraphics[height=6.5cm,clip=true]{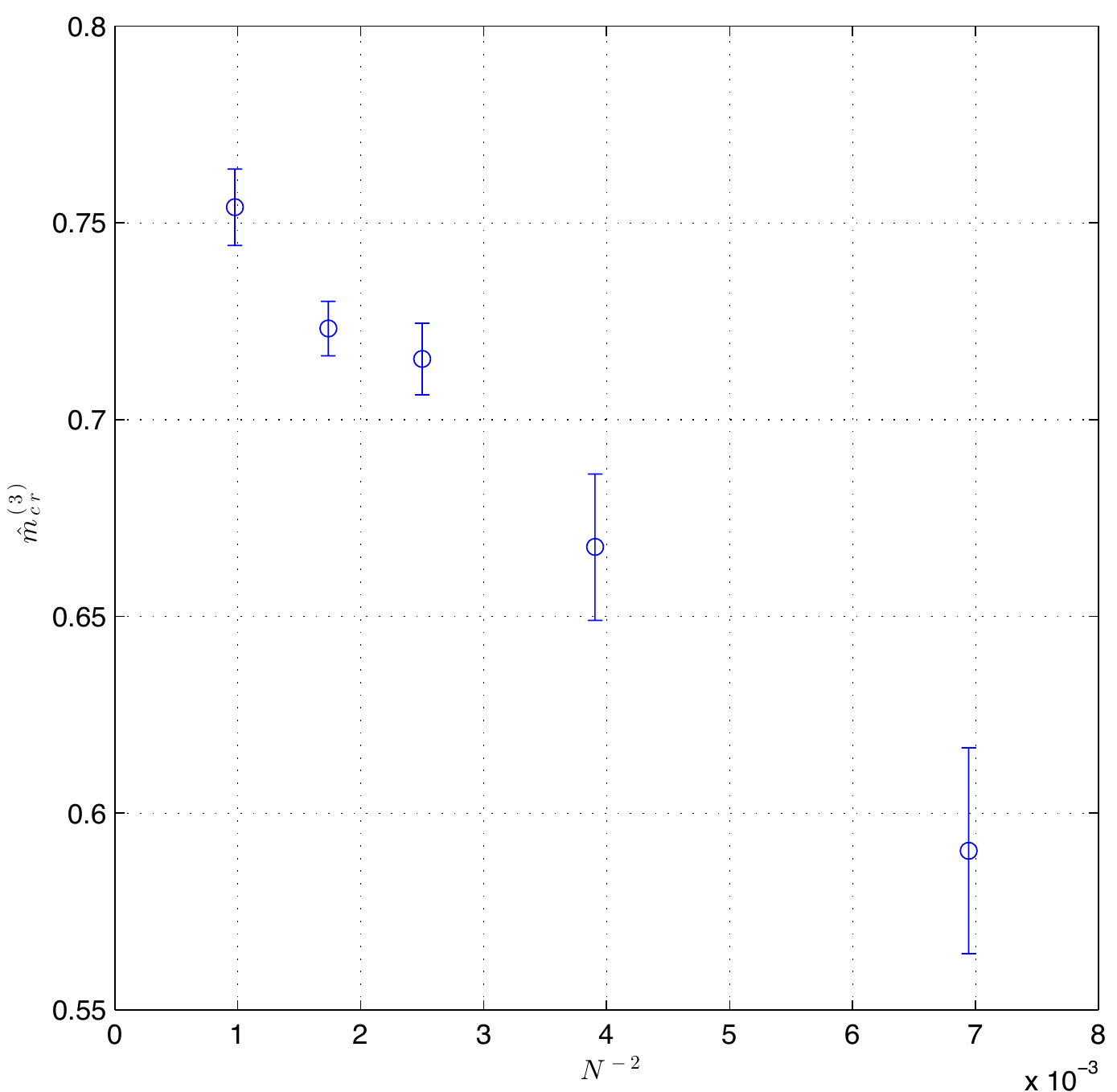}
  \end{tabular}
\end{center}
  \caption{Three-loop critical mass, for the $n_f=4$ case: zero momentum extrapolation 
on a $32^4$ lattice (left) and infinite volume extrapolation
(right). In the left panel, data (blue circles, with errorbars) and
fit results (red
circles, no errorbars) are plotted on top of each other.}
\end{figure}

\subsection{Fitting irrelevant and finite volume effects}

In Eq.~(\ref{eq:masterZ}) currents renormalization constants 
are defined in terms of the quark field renormalization
constant. The latter can be computed - see Eq.~(\ref{eq:Zq}) - from 
the quark self-energy, more precisely from its component along the
gamma matrices, which at any finite value of the lattice spacing reads
\begin{eqnarray}
\hat{\Sigma}_{\gamma} = \frac{1}{4} \sum_{\nu} \gamma_{\nu} {\rm Tr_{spin}}(\gamma_{\nu} \hat{\Sigma} ) 
& = & i\sum_\nu\gamma_\nu\hat p_\nu\left( 
\hat\Sigma_{\gamma}^{(0)}(\hat p)+ \hat
p^2_\nu\hat\Sigma_{\gamma}^{(1)}(\hat p)+ \hat
p^4_\nu\hat\Sigma_{\gamma}^{(2)}(\hat p) + \ldots \right) \nonumber \\
& \equiv & i\sum_\nu\gamma_\nu\hat p_\nu \; \widehat\Sigma_{\gamma}(\hat p, \nu)
\label{eq:sigmHAT}
\end{eqnarray}
Notice the tower of irrelevant contributions which go on top of the
one expected in the continuum limit. All these contributions are
contained in the definition of $\widehat\Sigma_{\gamma}(\hat p,
\nu)$. In the latter one recognizes a dependence on the direction 
$\nu$, which comes via the dependence on the length $|\hat
p_{\nu}|$. In the continuum limit this dependence drops out and, 
once one subtracts the logarithmic contribution discussed in 
subsection~\ref{sec:3loopZ}.1, 
the value of the finite part of $Z_q(\mu=p)$ is given by 
$\lim_{a\rightarrow 0}\hat\Sigma_{\gamma}^{(0)}(\hat p)$. This
observation on the dependence on $\nu$ of
$\widehat\Sigma_{\gamma}(\hat p,\nu)$ becoming immaterial 
in the continuum limit has to be be born in mind also in
the following, \eg in Eq.~(\ref{eq:ZoHAT}). 

The second ingredient in Eq.~(\ref{eq:masterZ}) is given by the 
quantities $O_{\Gamma}$. These have their lattice counterparts 
\[
\hat O_{\Gamma}(\hat p) = Tr\left(\hat P_{O_\Gamma} \hat
  \Gamma_\Gamma(\hat p)\right)
\]
For the vector and axial currents we eliminate 
dependences on directions like the one we have just discussed
in the case of $\widehat\Sigma_{\gamma}(\hat p,\nu)$, \eg
\[
\hat O_{\gamma}(\hat p) = \frac{1}{4} \sum_{\nu} Tr\left(\gamma_{\nu} \,\hat
  \Gamma_{\gamma_{\nu}}(\hat p)\right)
\]
The reason for getting rid of this dependence in this case while
retaining it in the case of $\widehat\Sigma_{\gamma}(\hat p,\nu)$ will
be clear in a moment. 

Our NSPT computations are performed on different, finite lattice sizes 
$N=L/a$ (our lattices are always isotropic); therefore one must expect 
finite size corrections. On dimensional grounds, we can expect a 
dependence on $pL$. Since we want to compute the currents renormalization
constants in both the continuum and the infinite volume limit, we need
to take two limits. This is done in the following form
\beq
\label{eq:ZoHAT}
Z_{O_\Gamma}(\mu=p,\beta^{-1}) |_{\mbox{\tiny{finite part}}}  = 
\lim_{\substack{a \rightarrow 0 \\ L \rightarrow \infty}} 
\widehat O_\Gamma(\hat p, pL, \nu) 
\equiv \lim_{\substack{a \rightarrow 0 \\ L \rightarrow \infty}} 
\frac{\widehat\Sigma_{\gamma}(\hat p, pL, \nu)}
{\hat O_{\Gamma}(\hat p, pL)} |_{\mbox{\tiny{log subtr}}}
\eeq

Eq.~(\ref{eq:ZoHAT}) is our key formula and deserves a few
comments: 
\begin{itemize}
\item We only compute the finite parts. As it was made clear in
  subsection~\ref{sec:3loopZ}.1, we know all the relevant logarithms
  entering the quantities we are concerned with. This means in
  particular that we can subtract their contribution: this is the
  meaning of the subscript $\ldots |_{\mbox{\tiny{log subtr}}}$ in the
  definition of $\widehat O_\Gamma(\hat p, pL, \nu)$.
\item Since in Eq.~(\ref{eq:ZoHAT}) we take the limits 
$a \rightarrow 0$ and $L \rightarrow \infty$ and we subtract 
the logs that come from the anomalous dimension 
$\gamma_{O_\Gamma}$, 
$\widehat\Sigma_{\gamma}(\hat p, pL, \nu)$ reconstructs the
contribution of $Z_q$ (and in this limit the dependence on
$\nu$ drops out). Notice that this is true because of the ratio that
is taken. In order to determine $Z_q$ itself one
should look at $\widehat\Sigma_{\gamma}(\hat p, pL, \nu)$ alone and
perform the subtraction of {\em different logs} (\ie, those connected
to the quark field anomalous dimension). 
\item On a fixed lattice volume, the $a \rightarrow 0$ 
limit of $\widehat O_\Gamma(\hat p, pL,\nu)$ can be evaluated 
by computing the quantity for different momenta 
$\hat p$ and fitting the results in terms of hypercubic invariants,
\eg those listed in Eq.~(\ref{eq:Hinv}). The possible terms are
dictated by symmetries of both $\widehat\Sigma_{\gamma}(\hat p, pL,
\nu)$ and $\hat O_{\Gamma}(\hat p, pL)$ (a formal power counting
fixes how many terms one should retain).
\item We want to account for the limits $a \rightarrow 0$ and 
$L \rightarrow \infty$ simultaneously. This is done by computing 
the quantity $\widehat O_\Gamma(\hat p, pL,\nu)$ on different volumes 
and performing a combined fit. The combined fit is made possible by
defining finite size corrections according to  
\begin{eqnarray}
\widehat O_{\Gamma}(\hat p, pL,\nu) & = & \widehat O_{\Gamma}(\hat p,
\infty,\nu) + \left( \widehat O_{\Gamma}(\hat p, pL,\nu) -
  \widehat O_{\Gamma}(\hat p, \infty,\nu) \right) \nonumber \\
& \equiv & \widehat O_{\Gamma}(\hat p, \infty,\nu) + 
\Delta\widehat O_{\Gamma}(\hat p, pL,\nu) \nonumber \\
& \simeq & \widehat O_{\Gamma}(\hat p, \infty,\nu) + 
\Delta\widehat O_{\Gamma}(pL)
\end{eqnarray}
where the main rationale for the last (approximate) equality is 
that {\em we neglect corrections on top of corrections}. Since 
$p_\mu L = \frac{2\pi n_\mu}{L}L = 2\pi n_\mu$, 
there is only one finite size correction for each 4-tuple
$\{n_\mu\,|\,\mu=1,2,3,4\}$ and no functional form has to be inferred for the
correction.
\end{itemize}
All in all, a prototypal fitting form of ours reads
\beq
\label{eq:FITexample}
\widehat O_{\Gamma}(\hat p, pL,\nu)  = 
c_1 +
c_2 \sum_\sigma \hat p_\sigma^2 +
c_3 \frac{\sum_\sigma \hat p_\sigma^4}{\sum_\rho \hat p_\rho^2} +
c_4 \hat p^2_{\nu} + 
\Delta\widehat O_{\Gamma}(pL) + \mathcal{O}(a^4).
\eeq
where in order to make things easy we limited to a very moderate order
in $a$ (actually, less than what we use in realistic fits). The relevant
contribution to the finite part is $c_1$. We recall once
again that this is a combined fit to data taken on different lattice
sizes, with the same $\Delta\widehat O_{\Gamma}(pL)$ applying to all
the data corresponding to the same momentum 4-tuple
$\{n_1,n_2,n_3,n_4\}$ (resulting in different 
values of momenta on different lattice sizes). Notice that the
inclusion of $\Delta\widehat O_{\Gamma}(pL)$ makes the fit not
constrained with respect to an overall shift. This is cured by
including in the fit a few measurements (of the order $\sim 1,2,3$) 
taken on the largest lattice in the high momentum region: these are
assumed free of finite size effects and act as a normalization point. 

Finite size effects can be easily spotted in the left panel of
Fig.~2, where we plot one-loop $\widehat O_S(\hat p, pL, \nu)$
on both $32^4$ (black symbols) and $16^4$ (red, filled symbols), in
the $n_f=4$ case. One
can see that data are arranged in {\em families} (different symbols):
this is a direct consequence of the dependence on $\nu$ one can see \eg
in Eq.~(\ref{eq:FITexample}). All in all, there is one family for each
length $|\hat p_{\nu}|$. Since finite size effects are there, families
do not join smoothly across different lattices. In the right panel of 
Fig.~2 one can inspect how things change taking into account finite
size corrections $\Delta\widehat O_{S}(pL)$: families do join
smoothly. This family mechanism provides a very effective handle to
detect finite size effects: this is the reason for retaining the $\nu$
dependence in the definition of $\widehat O_{\Gamma}(\hat p, pL,\nu)$;
to be definite, we retain it in the numerator (\ie, in the
contribution connected to $Z_q$); keeping it also in the denominator
would result in a too odd fitting form.

\begin{figure}[!t]
\begin{center}
  \begin{tabular}{cc}
     \includegraphics[height=6.5cm,clip=true]{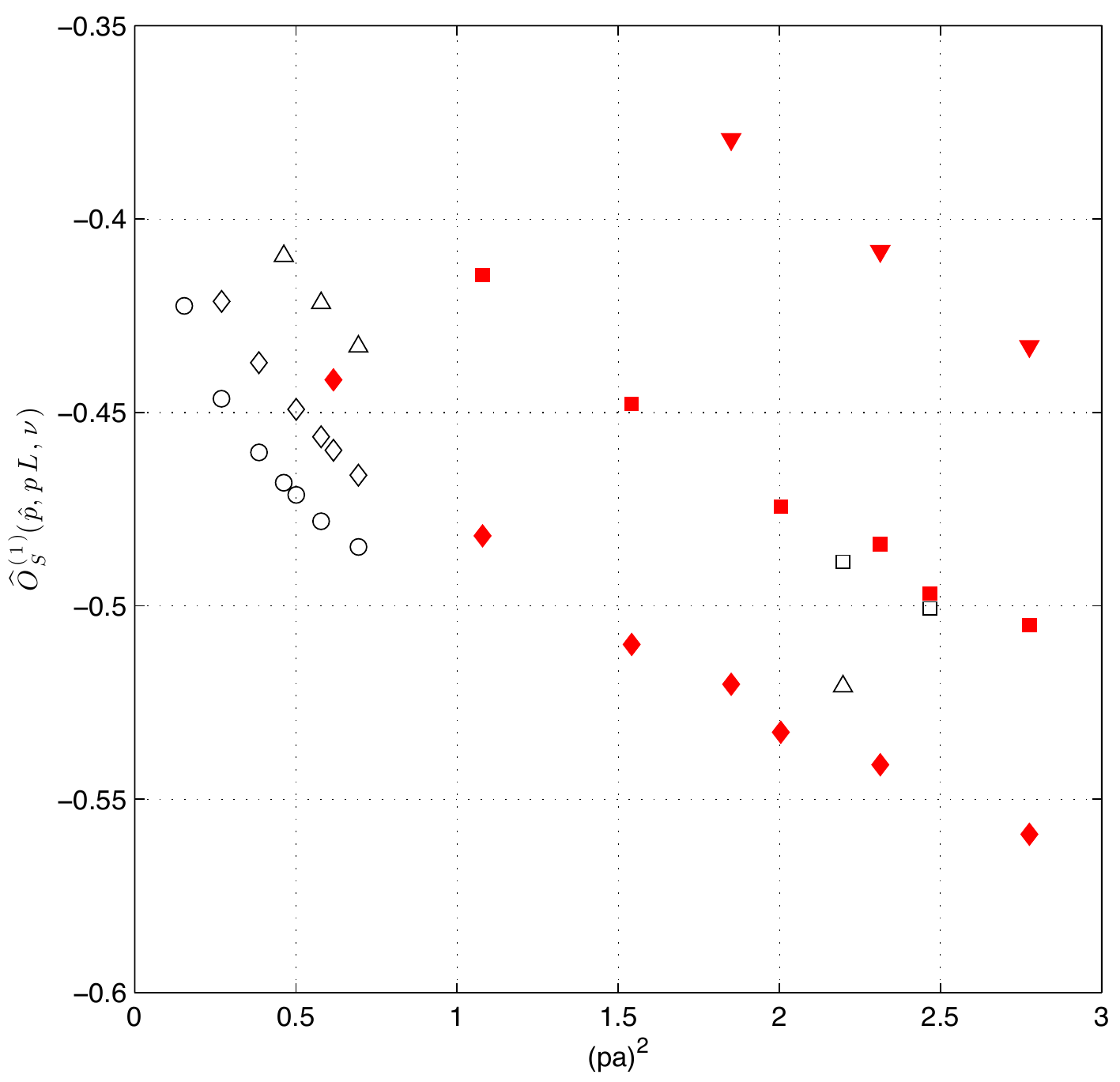}
     &
 \hspace{1.cm}
     \includegraphics[height=6.5cm,clip=true]{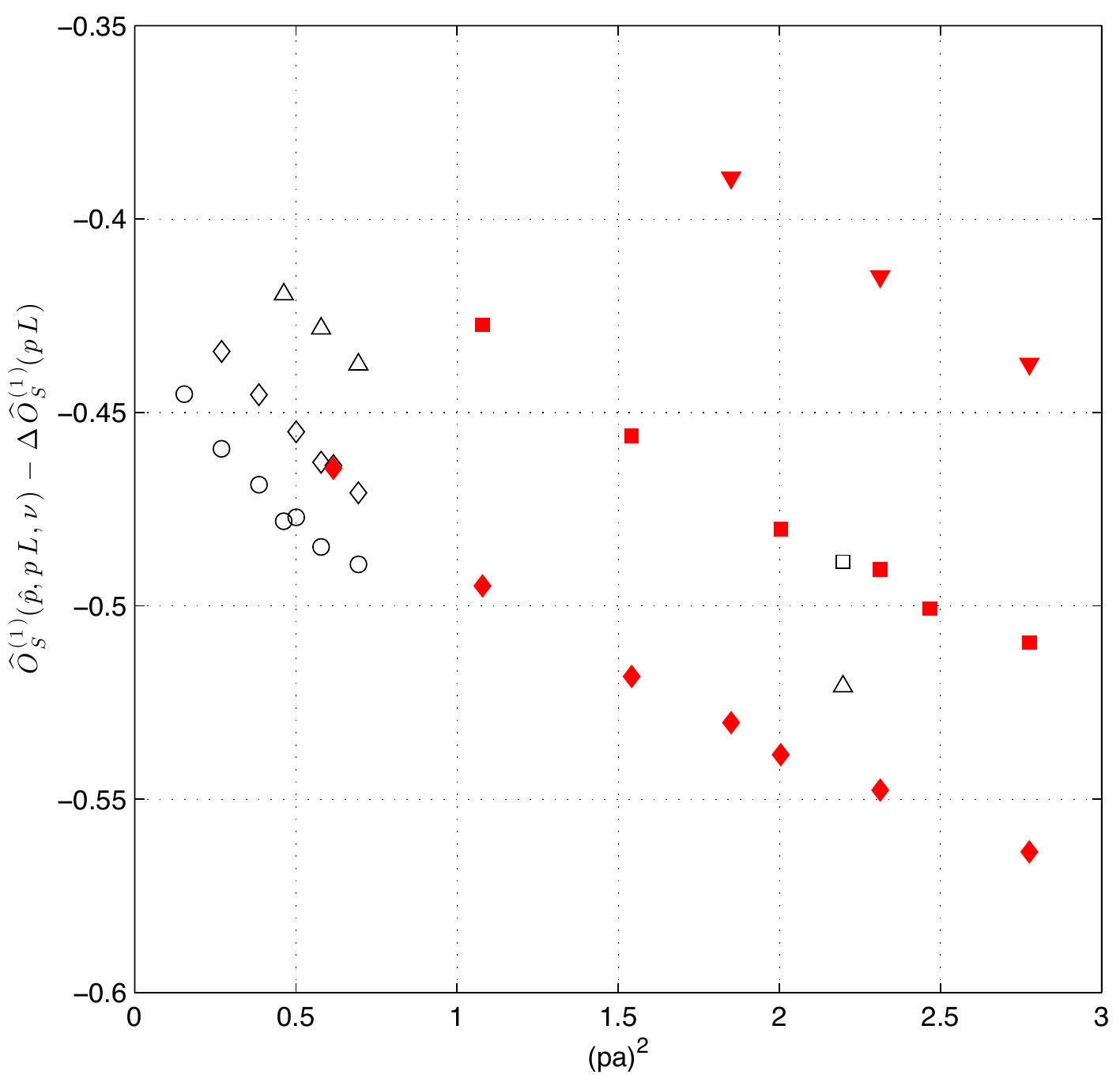}
  \end{tabular}
\end{center}
  \caption{One-loop ($n_f=4$) $\widehat O_S(\hat p, pL, \nu)$ (see Eq.~(15))
    measured on a $32^4$
    (black, empty) and a $16^4$ (red, filled) lattice,  
    without (left) and with (right) finite size corrections. 
Notice that, in the right panel, the two (black/empty and red/filled) square
data points near
$(pa)^2=2.5$ literally fall on top of each other (as they should, if
finite size effects were indeed perfectly removed), so that only 
one can be seen.}
\end{figure}

\section{Results}
\label{sec:res}

In Table~1 we give a brief account of our statistics. One can see that
computations were performed on five different sizes ($32^4$, $20^4$, 
$24^4$, $16^4$, $12^4$) for both $n_f=0$ and $n_f=4$. The latter are
our main point of interest. While with only two
values of $n_f$ we can not determine the coefficients of the $n_f$ 
dependence\footnote{At two-loop one could of course pin down a number, 
but that would not even have the status of a fit.}, it is
interesting to have at least an indication of how sensitive results are to
the number of flavors. 

Since NSPT
requires the (order by order) integration of the Langevin equation, a
finite order integration scheme for this stochastic differential
equation is needed: the Euler scheme is our choice here. An 
$\epsilon \rightarrow 0$ extrapolation (a linear one, in the
case at hand)  is needed to remove the effects of the finite time step
$\epsilon$. In Table~1 we provide the statistics collected on each
different size. Notice that the values of $\epsilon$ are chosen
different for $n_f=0$ and $n_f=4$ (see the discussion in
\cite{NSPT1}). 

Configurations were saved on which we can still
measure different observables. They were saved with frequencies 
which were chosen having a rough analysis of the autocorrelations in
place. The analysis of the different observables is in charge of
dealing with the residual autocorrelation effects.

\begin{table}[t]
\caption{Number of measurements at different 
value of the time step for the different lattice sizes, for both $n_f=4$
and $n_f=0$.}
\begin{center}
\begin{tabular}{|c|c|c|c|c|c|c|}
\hline
lattice size & $n_f=4$ & $n_f=4$ & $n_f=4$ & $n_f=0$ & $n_f=0$ & $n_f=4$ \\
$N = L/a$  & $\epsilon = 0.005$ & $\epsilon = 0.010$ & $\epsilon = 0.015$ & $\epsilon = 0.010$ & $\epsilon = 0.020$ & $\epsilon = 0.030$ \\
\hline
\hline
$12$ & 210                       & 210 & 208  & 220                       & 209 & 209 \\
$16$ & 179                       & 185 & 148 & 111                       & 119 & 116 \\
$20$ & 84                    & 84 & 82   & 71                    & 70 & 70 \\
$24$ & 72                    & 74 & 79 & 54                    & 54 & 52 \\
$32$ & 49                    & 47 & 49 & 42                    & 46 & 45 \\
\hline
\hline
\end{tabular}\\
\end{center}
\end{table}
In Table~2 we report the coefficients of the three-loop expansion of 
$Z_S$, $Z_P$, $Z_V$ and $Z_A$\footnote{The reader will notice a few 
significant corrections with respect to the preliminary 
results in \cite{NSPT_Zs_LAT2012}.}. Results are
reported for both $n_f=4$ and $n_f=0$.
We remind the reader that the expansion parameter is $\beta^{-1}$.
We stopped at three-loop order given the knowledge of
anomalous dimensions which we can get from \cite{Gracey}. 
For  finite quantities there is in principle no limitation 
(other than practical ones dictated by
statistics). Notice that \cite{GraceyNEW} could 
now open new opportunities for even higher order computations. 
\begin{table}[!b]
\caption{One-, two- and three-loop coefficients of the renormalization 
constants for quark bilinears for both $n_f=4$ and $n_f=0$. Expansions are in
$\beta^{-1}$. One-loop analytical results are reported for comparison.}
\begin{center}
\begin{tabular}{|c|c|c|c|c|c|c|}
\hline
\, & \small {\em analytical} & \, & $n_f=4$ & $n_f=0$ & $n_f=4$ & $n_f=0$\\ 
\,  & \small {\em one-loop} & one-loop & two-loop & two-loop & three-loop & three-loop \\
\hline
\hline
$Z_S$ & -0.4488 & -0.442(6) & -0.170(11) & -0.228(9) & -0.33(11) & -0.39(11)\\
$Z_P$ & -0.7433 & -0.739(7) & -0.202(13) & -0.309(11) & -0.58(11) & -0.71(12)\\
$Z_V$ & -0.5623 & -0.561(7) & -0.067(12) & -0.147(9) & -0.367(61) & -0.463(55)\\
$Z_A$ & -0.4150 & -0.419(6) & -0.033(12) & -0.097(8) & -0.236(56) & -0.299(48)\\
\hline
\hline
\end{tabular}
\end{center}
\end{table}
We quote the analytical one-loop results \cite{Aoki}: the 
comparison is a first proof of the effectiveness of our method. 
The reader could notice that we have somehow less systematic 
deviations from analytic results here than in \cite{NSPT_TLS}. 
This is due to the fact that we took the normalization points for
finite size effects at slightly higher values of 
momentum: see the discussion after Eq.~(\ref{eq:FITexample}).
To be definite, the normalization points are in the highest 
$(pa)^2$ region of Figure~2, which in terms of n-tuple reads 
$\{4,4,4,4\}, \{4,4,4,5\}$; the two choices have been compared and 
results have been proved to be equivalent within errors. 
A more
stringent confirmation of our results comes from the fitting of
irrelevant contributions. The latter is a key ingredient of our
approach, since fitting irrelevant contributions compliant to lattice
symmetries is our handle on continuum limit. A comparison to results
in \cite{Haris} made us confident in our results: the leading
irrelevant terms that we fitted are consistent with the diagrammatic
results. 

The errors we quote are dominated by the stability of fits with respect to the
change of fitting ranges, functional forms, number of lattice sizes
simultaneously taken into account. Notice that three-loop results
for $Z_S$ and $Z_P$ have an extra source of error in the
indetermination in the coupling matching parameter $X$ (see Eq.~(\ref{eq:ourX}) 
and the discussion over there). 

All in all, our new results, \ie $\,$ two-loop and three-loop contributions to
$Z_S$, $Z_P$, $Z_V$ and $Z_A$ for Iwasaki action
seem to be quite moderate, in particular two-loop: taking into account
the typical values of $\beta$ that are relevant to numerical
simulations ($\beta \sim 2$), three-loop contributions are typically
larger than two-loop. Coefficients themselves are 
smaller than the ones found in the case of tree-level Symanzik 
improved action \cite{NSPT_TLS}, but this is not {\em per se} 
any significant. First of all, the comparison in magnitude of two- 
and three-loop coefficients has to be corrected for the different 
$\beta$ value regimes one is interested in (there is roughly a factor of $2$). 
What is even more important is the weight of two- and three-loop
contributions with respect to the leading one: 
what really makes the difference in between the two different
regularizations is the relative weight of one-loop contributions themselves.   

Another general feature that emerge from our computations is that 
irrelevant corrections from hypercubic invariants which are not $O(4)$ invariant
appear to be in general quite significant for this action. All this is of
course a numerical accident, but it is a relevant one when it comes
to summing the series and assessing irrelevant effects.

\section{Summing the series}
\label{sec:sum0}

We now come to the issue of summing the series. In this and in the
following Section we will deal not only with the results of this paper,
but also with the ones for the regularization defined by tree-level
Symanzik improved gauge action and $n_f=2$ Wilson fermions, \ie
$\,$ those of \cite{NSPT_TLS}.

For both cases one can compare perturbative and non-perturbative
results, which for Iwasaki can be found in \cite{ETMC_ZsIWA}. 
As already said, \cite{ETMC_ZsIWA} deals with the same, massless RI-MOM
scheme with Twisted Mass fermions: results are presented at
$\beta=1.95$ and $\beta=2.10$. 

For $Z_V$ we obtain 
\[
Z_V(\beta=1.95) = 0.644(11)(49) \;\;\;\;\; 
Z_V(\beta=2.10) = 0.677(9)(39), 
\]
where the first error is the
statistical one, while the second is a rough estimate of the
truncation effects, which we simply take as the highest
order contribution. The latter recipe is the conventional one: one is
of course well aware of its roughness, even if making use of it at
three-loop level is more than what is usually done. 
On the other side, we have already made the point that two-loop contributions
are indeed small for the Iwasaki case (and indeed even smaller for the 
$n_f=4$ than for the $n_f=0$ case). Since three-loop contribution is
thus relatively important, the net effect is an estimate of
truncation errors which is quite large. 
The other finite renormalization constant is
$Z_A$, for which we obtain 
\[
Z_A(\beta=1.95) = 0.747(11)(32)  \;\;\;\;\; 
Z_A(\beta=2.10) = 0.769(9)(25). 
\]
For $Z_S$ we get in turn 
\[
Z_S(\beta=1.95) = 0.681(18)(44) \;\;\;\;\; 
Z_S(\beta=2.10) =0.712(14)(36).
\] 
Finally $Z_P$ reads 
\[
Z_P(\beta=1.95) = 0.487(18)(78) \;\;\;\;\; 
Z_P(\beta=2.10) = 0.538(15)(63).
\]

One can directly inspect a fair agreement with the results of
\cite{ETMC_ZsIWA}: basically the errors that result from our procedure
make the perturbative and non-perturbative results fully consistent. 
Actually the (smaller) statistical errors
would be enough to obtain a substantial agreement with
non-perturbative 
results in the case of the finite renormalization constants. Notice
that non-perturbative results in \cite{ETMC_ZsIWA} are presented in two
variants, referring to the different prescriptions ``M1'' and ``M2''
which the authors discuss. Here it suffices to say that the
differences are to be ascribed to different treatments of irrelevant 
effects (and thus they are  systematic effects): in general the two
methods differ more for the divergent than for the
finite renormalization constants. We will have more to say on
irrelevant effects later in this Section and then again 
in Section \ref{sec:whatNEW}.

We now move to the results we got for the regularization defined by
tree-level Symanzik improved gauge action and $n_f=2$ Wilson fermions 
(\ie $\,$ those in \cite{NSPT_TLS}). We stress once again that in this
case the two-
and three-loop coefficients are larger, but this should be corrected
by taking into account the regime of $\beta$ one is interested in
(typical values of $\beta$ for tree-level Symanzik are roughly double
of those for Iwasaki). Moreover, convergence properties of the series
are dominated by the relative weights of one-loop and higher orders
contributions. The results we obtain summing the series we
computed in \cite{NSPT_TLS} can be compared
to the non-perturbative ones in \cite{ETMC_Zs}. In this case we make
our comparison at the largest value of $\beta$ which is discussed in
\cite{ETMC_Zs} (the reason for this will be clear in a moment). For
$Z_V$ our results sum to 
\[
Z_V(\beta=4.05) = 0.710(2)(28) 
\]
while 
\[
Z_A(\beta=4.05) = 0.788(2)(18). 
\]
Moving to logarithmically divergent renormalization constants, we get 
\[
Z_S(\beta=4.05) = 0.753(4)(30) 
\]
and\footnote{We regret a typo in the value of $Z_P$ reported in 
\cite{NSPT_TLS}. }
\[
Z_P(\beta=4.05) = 0.601(5)(48).
\]
Conventions with errors are the same as before. In this case deviations
are manifest, in particular for $Z_S$ and $Z_P$. This in the end does not come as a surprise, given the
observations we have already made: convergence properties
 are strongly controlled by the relative weight of one-loop and
higher-order contributions.
This is the reason for not attempting to sum the series at
values of $\beta$ smaller than the largest one. While there is a
tendency to converge for finite quantities, logarithmically divergent
constants are fairly away from each other in the perturbative and
non-perturbative computations. This clearly motivates the step forward
of summing the series in different couplings, which will be addressed
in the following Section. Before we move to that issue, we present a
first discussion of how we can assess the impact of irrelevant effects once we
sum the series.

\begin{figure}[!t]
\begin{center}
  \begin{tabular}{cc}
     \includegraphics[height=6.5cm,clip=true]{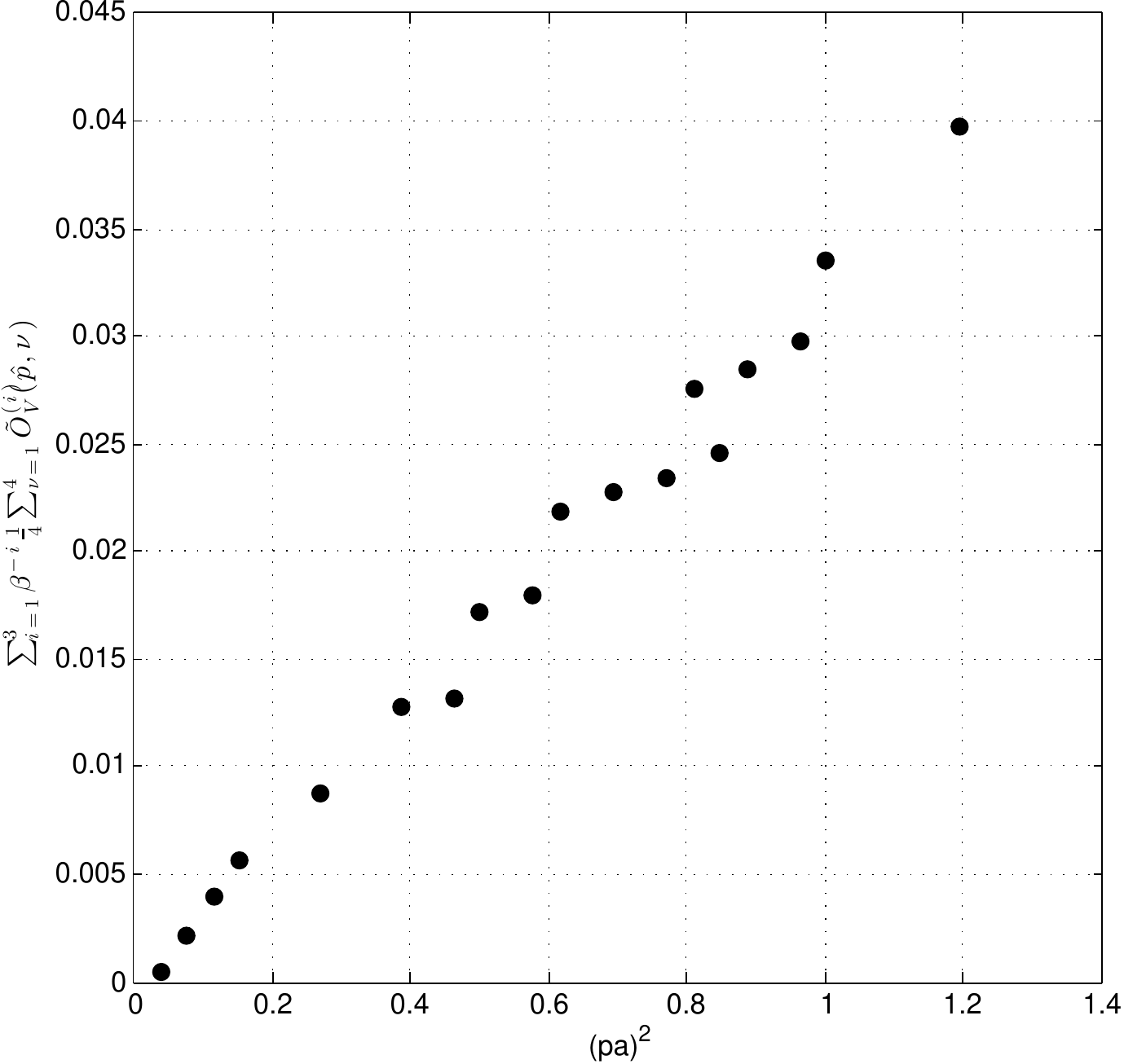}
     &
 \hspace{1.cm}
     \includegraphics[height=6.5cm,clip=true]{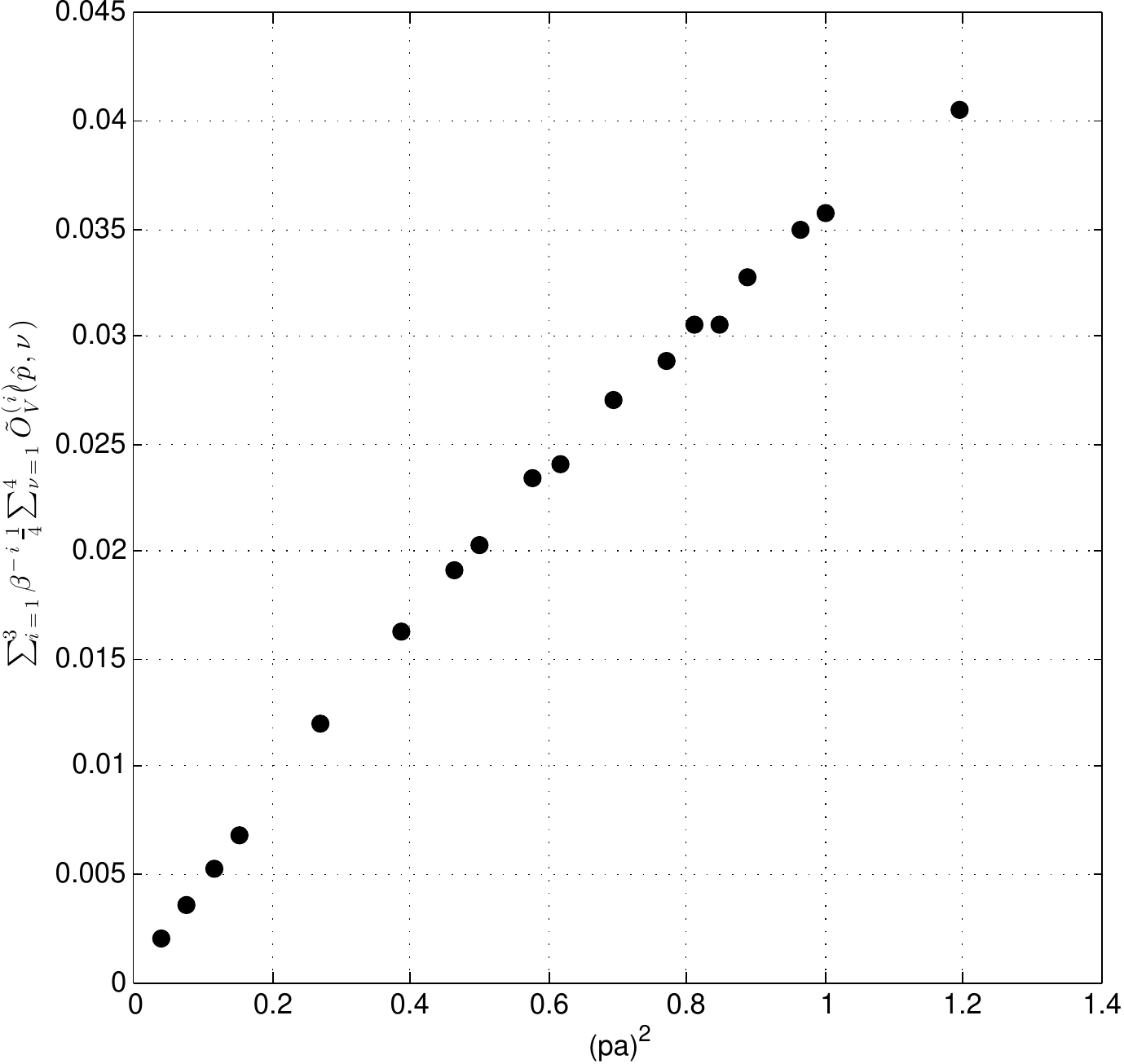}
  \end{tabular}
\end{center}
  \caption{The quantity $\sum_{i=1}^3 \beta^{-i} \, \frac{1}{4} \sum_{\nu=1}^4 
\tilde O_V^{(i)}(\hat p,\nu)$ for the Iwasaki (left; $\beta=2.10$) and Symanzik
(right; $\beta=4.05$) case. Black points quantify the impact of
irrelevant (finite lattice spacing) effects on a determination of $Z_V$.}
\end{figure}
 
The results we have just reported holds in the continuum and
infinite volume limits, {\em i.e.} they are free from irrelevant
and finite size effects. To be definite: in the prototypal
form of Eq.~(\ref{eq:FITexample}) this corresponds to retaining only $c_1$.
On the other side, to assess the irrelevant effects we can discard 
the continuum limit and finite-size contributions. Again, in the prototypal
form of Eq.~(\ref{eq:FITexample}) this corresponds to discarding 
$c_1$ (the continuum limit result) and $\Delta\widehat O_{\Gamma}(pL)$ 
(the finite size effects). This defines a new quantity, which we
denote $\tilde O_{\Gamma}(\hat p, \nu)$. At the same (very) moderate order 
of Eq.~(\ref{eq:FITexample}) a prototypal form for this quantity reads
\beq
\label{eq:irrelFITexample}
\tilde O_{\Gamma}(\hat p, \nu)  = 
c_2 \sum_\sigma \hat p_\sigma^2 +
c_3 \frac{\sum_\sigma \hat p_\sigma^4}{\sum_\rho \hat p_\rho^2} +
c_4 \hat p^2_{\nu} + \mathcal{O}(a^4).
\eeq
All in all: in $\tilde O_{\Gamma}(\hat p, \nu)$ everything depends on
(powers of) $\hat p$ and thus does not survive the continuum limit; on
the other side, there is no $pL$ dependence because that has been eliminated by
subtracting the $\Delta \widehat O_V(pL)$. Obviously, for divergent
constants we compute the finite parts only ({\em i.e.} these
are log-subtracted quantities).

In Figure~3 we plot the quantity
\[
\sum_{i=1}^3 \beta^{-i} \, \frac{1}{4} \sum_{\nu=1}^4 
\tilde O_V^{(i)}(\hat p,\nu)
\]
for the Iwasaki (left panel) and the Symanzik (right panel) case 
(values of the coupling are once again 
$\beta=2.10$ and $\beta=4.05$ respectively). These can be regarded 
as the irrelevant contributions to $Z_V$ (computed in infinite volume
at three-loop accuracy). Notice that in abscissa we report
values of momentum in dimensionless units (in other terms, there is no
value for the lattice spacing involved). Notice also that in Figure~3
we average on directions, which is the common practice. 
When computed in this way, irrelevant
effects come out of our fit, which is necessarily an effective one: we
have to stop at a given order in the lattice spacing. We stress
nevertheless that the fit is performed at fairly large orders
(typically $a^6$) and at three-loop level.

One can easily see how different is the impact of violations of
(continuum-like) rotational symmetry in the two cases. It is true that
one often tries to minimize these effects by a convenient choice of
the momenta. One should nevertheless keep in mind the trivial
observation that the amount of
violation is not decided by the choice of momenta: one should try in
any case to fit terms compliant to the lattice symmetries. 

In Figure~4 we plot (in the case of Symanzik at $\beta=4.05$) the
observable relevant for computing $Z_V$, in yet another couple of
ways. Let's consider once again the prototypal expansion of 
Eq.~(\ref{eq:FITexample}) and let's define two other quantities that
at the same (moderate) order read
\begin{eqnarray}
\bar O_{\Gamma}(\hat p, \nu) & = & c_1 +
c_2 \sum_\sigma \hat p_\sigma^2 +
c_3 \frac{\sum_\sigma \hat p_\sigma^4}{\sum_\rho \hat p_\rho^2} +
c_4 \hat p^2_{\nu} + \mathcal{O}(a^4) \nonumber \\
\mathring O_{\Gamma}(\hat p) & = & c_1 +
c_2 \sum_\sigma \hat p_\sigma^2 +
c_3 \frac{\sum_\sigma \hat p_\sigma^4}{\sum_\rho \hat p_\rho^2} +
\mathcal{O}(a^4) 
\end{eqnarray}
All in all: out of the fit results, in both cases we discard the
finite size contributions and in the second quantity we also discard
what depends on the length $|\hat p_{\nu}|$, {\em i.e.} we cut part of
the irrelevant effects\footnote{One can see a more general effect of 
the recipe for $\mathring O_{\Gamma}(\hat p)$ by referring to 
Eq.~(\ref{eq:sigmHAT}): over there the recipe amounts to singling out 
the contribution of the $\hat\Sigma_{\gamma}^{(0)}(\hat p)$ term.}.

\begin{figure}[!b]
\begin{center}
  \begin{tabular}{cc}
     \includegraphics[height=6.5cm,clip=true,angle=90]{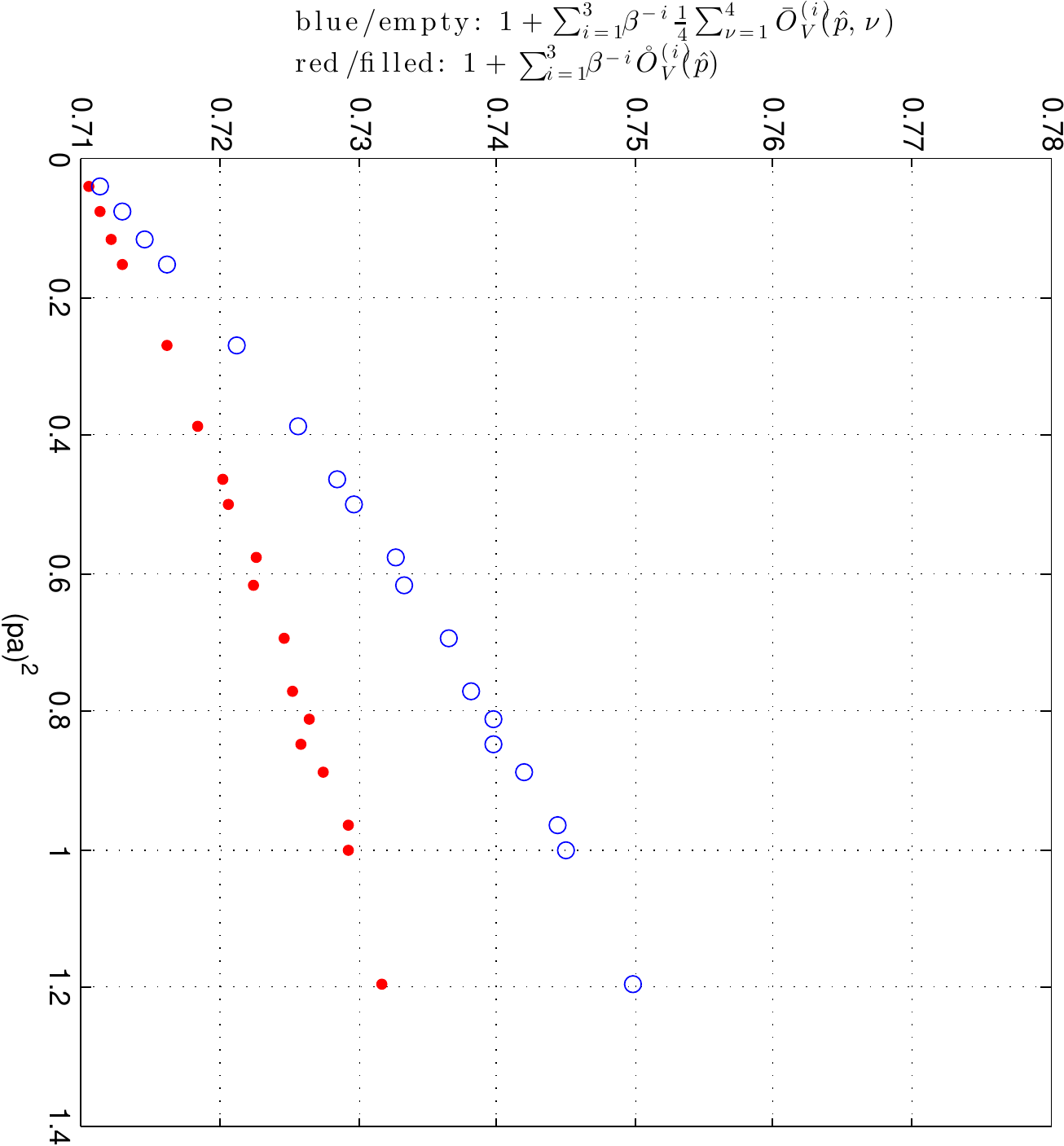}
     &
 \hspace{1.cm}
     \includegraphics[height=6.5cm,clip=true,angle=90]{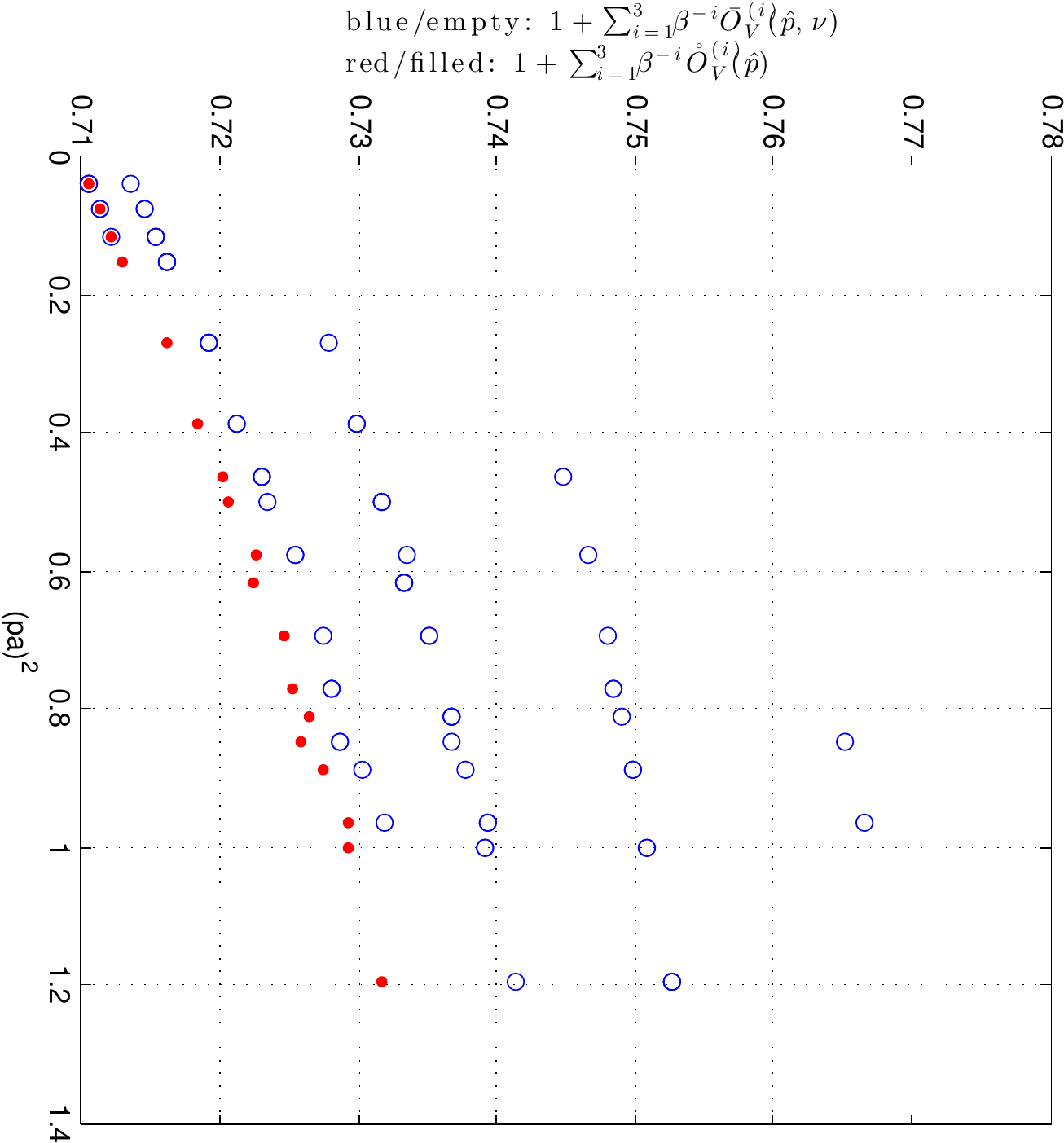}
  \end{tabular}
\end{center}
  \caption{On both panels, the red/filled circles mark the quantity 
$1+\sum_{i=1}^3 \beta^{-i} \, \mathring O_V^{(i)}(\hat p)$. As for 
blue/empty circles: on the left panel they denote the quantity 
$1+\sum_{i=1}^3 \beta^{-i} \,  \, \frac{1}{4} \sum_{\nu=1}^4 
\, \bar O_V^{(i)}(\hat p,\nu)$, while on the right they are 
$1+\sum_{i=1}^3 \beta^{-i} \, \bar O_V^{(i)}(\hat p,\nu)$. Data are 
for Symanzik action at $\beta=4.05$.}
\end{figure}
 
On the left panel of Figure~4 blue/empty circles denote the
quantity 
\[
1+\sum_{i=1}^3 \beta^{-i} \,  \, \frac{1}{4} \sum_{\nu=1}^4 
\, \bar O_V^{(i)}(\hat p,\nu), 
\]
which is averaged over
directions. Again on the left panel, red/filled circles denote instead 
\[
1+\sum_{i=1}^3 \beta^{-i} \, \mathring O_V^{(i)}(\hat p).
\]

On the right panel, the red/filled circles are the same as on 
the left, while the blue/empty circles denote instead the
quantity 
\[
1+\sum_{i=1}^3 \beta^{-i} \, \bar O_V^{(i)}(\hat p,\nu), 
\]
for which there is no average over directions and {\em families} 
come into place again. There is one subtlety: one can see
that in the right panel the blue, empty circles do not point 
{\em in a trivial way}
to the result one is interested in. In other terms, when we keep the
{\em families} structure, only the red, filled
circles are the ones smoothly guiding the eye to the correct extrapolated
result.

\section{Summing the series in different couplings}
\label{sec:sum}

The Symanzik case displayed not so brilliant convergence properties. 
Thus, that is the prototypal situation in which one would like
to go for what is usually, generically referred to as Boosted
Perturbation Theory \cite{LepMack}. One re-expresses the series as
expansions in different couplings, of course looking for better
convergence properties that in the case one starts
with. Often one deals with this having only a one-loop result
available. As discussed in \cite{NSPT_Zs}, this is at risk of being an
empty exercise. At one loop, nothing changes but the value of the
coupling itself. So, the effectiveness of the procedure relies on
the optimal choice of coupling and scale that are really relevant
for the computation at hand. Actually this choice has to be regarded
as so good that one-loop captures essentially the complete result. 
This does not need to hold true and can be strictly speaking only 
assessed {\em a posteriori}. 
Only having at least a two-loop result available one
can inspect how the series actually reshuffle and one can hope to
learn something more on the convergence properties. Our question is: 
can a three-loop computation be reliable enough to gain solid, new
pieces of information? 

We here compare results obtained as expansions in the couplings 
which were also used in \cite{NSPT_Zs}, \ie
\[
x_0 = \beta^{-1} \;\;\;\; x_1 \equiv \frac{\beta^{-1}}{\sqrt{P}}  
\;\;\;\; x_2 \equiv - \frac{1}{2} \log(P)  \;\;\;\; x_3 \equiv \frac{\beta^{-1}}{P} \,.
\]
$P$ is the basic $1\times 1$ plaquette, for which we do have an
expansion in $\beta^{-1}$. In the Symanzik, $n_f=2$ case, the latter reads 
\[
P^{(TLS,n_f=2)} = 1 - 1.4649(12) \beta^{-1}  -0.2730(7) \beta^{-2} -0.6536(18)
\beta^{-3} + \ldots.
\]
For the Iwasaki, $n_f=4$ case, we have\footnote{By a mere numerical
  accident, in this case the error on the three-loop coefficient is actually smaller
  than that on the two-loop coefficient.}
\[
P^{(Iwa,n_f=4)} = 1 -0.8410(1) \beta^{-1}  +0.1328(63) \beta^{-2} -0.2014(4) 
\beta^{-3} + \ldots.
\]
We can thus work out the expansions we are interested in. 
$x_2$ and $x_3$ are quite popular as boosted couplings. 
In the end, we want to see whether results coming from summing series in different
couplings do or do not all approach the same result. The definition of 
$x_1$ can be useful with this respect. Convergence properties in 
the Iwasaki computations are fairly good in the original coupling; 
we will focus the case of Symanzik action, looking for better
convergence. There
is an overall ambiguity we have to live with: we do not have
non-perturbative simulations in the same setting we are dealing with
(Symanzik action and $n_f=2$ Wilson fermions). In view of this
limitation, we have no non-perturbative value for the different couplings. 
We have indeed estimates which come in turn from summing perturbative 
expansions of the plaquette (actually even at higher orders than
three-loop): these are the values we plug in. On the other side, one
could even take in first approximation the values of the plaquette for 
the different regularization of \cite{NSPT_Zs}. This ambiguity is admittedly a
limitation. Still, if we take into account the order of magnitude of the error one can attach to
the value of the coupling, it turns out that this is dominated by the
other errors, typically the truncation errors which are still the
dominant ones. The latter are estimated as done previously (\ie, as
the highest order contribution) and will be the only ones reported in
the following. Table 3 summarizes our results. 

\begin{table}[!t]
\caption{Quark bilinears renormalization constants for 
tree-level Symanzik improved gauge action and $n_f=2$ Wilson fermions,
at $\beta=4.05$, summed in different couplings. Non-perturbative results from
ref.\cite{ETMC_Zs} are reported for comparison.}
\begin{center}
\begin{tabular}{|c|c|c|c|c|c|c|}
\hline
\, & \small {expansion in} & \small {expansion in} & \small {expansion in} & \small {expansion in} & ref.\cite{ETMC_Zs} & ref.\cite{ETMC_Zs}\\ 
\,  & $x_0 = \beta^{-1}$ & $x_1 \equiv \frac{\beta^{-1}}{\sqrt{P}} $ & $x_2 \equiv - \frac{1}{2} \log(P)$ & $x_3 \equiv \frac{\beta^{-1}}{P}$ & (M1) & (M2) \\
\hline
\hline
$Z_V$ & 0.710(2)(28) & 0.686(21) & 0.688(17) & 0.661(55) & 0.659(4) & 0.662(3)\\
$Z_A$ & 0.788(2)(18) & 0.773(12) & 0.775(9) & 0.763(26) & 0.772(6) & 0.758(4)\\
$Z_S$ & 0.753(4)(30) & 0.727(29) & 0.726(27) & 0.705(49) & 0.645(6) & 0.678(4)\\
$Z_P$ & 0.601(5)(48) & 0.558(45) & 0.558(41) & 0.526(73) & 0.440(6) & 0.480(4)\\
\hline
\hline
\end{tabular}
\end{center}
\end{table}

Let's start from looking at $Z_V$. Notice that switching from $x_0$ to 
$x_1$ and then to $x_2$, the value of the couplings are getting larger and
larger as we proceed. Results for the $x_1$ and $x_2$ expansions 
are quite close to each other and
they both approach the results of \cite{ETMC_Zs}. We get even closer
when we switch to $x_3$.
While the central value is now
literally on top of the non-perturbative result, the error has
become pretty large. This is simply the effect of the fact that the 
series has started oscillating: already at one loop one gets
essentially the result $0.66$, and
then two- and three-loop contributions basically cancel each other. 

We proceed to $Z_A$. 
Once again, in the case of the $x_3$ expansion the series has already started
oscillating. All in all, it is fair to say that results for finite
constants display a tendency to get closer to
the non-perturbative ones. Actually, $Z_V$ changed more than $Z_A$,
which is good, since the former was deviating more than the latter 
from non-perturbative results. 

We proceed to the logarithmically divergent renormalization
constants. If one takes the values of $Z_S$ and $Z_P$ after the (various,
different) boosting procedures and compare them to the results 
in \cite{ETMC_Zs}, one can still see quite
important discrepancies. So, there is still quite a gap for divergent 
renormalization constants, which
did not hold true in the case of finite constants. It could well be
that one simply needs more terms to definitely assess the convergence
properties, but there is another issue which could be considered. We
have already noticed that ``M1'' and ``M2'' results in \cite{ETMC_Zs}
differ much more in the case of $Z_S$ and $Z_P$ than in the case 
of $Z_A$ and $Z_V$. One method tries to gain more information from
the lower momenta region than the other. To be more precise, one
simply subtracts the
known leading one-loop $a^2$ irrelevant effects and look for a
plateaux region, while the other tries to fit extra irrelevant effects
in the lower momenta region. This region is just the theater of  
a subtle interplay of UV and IR effects: from one side higher powers
of $pa$ are suppressed (and this is good to assess irrelevant 
contributions), but from another side that is just the region which
is prone to suffer from finite size (IR) effects. 

All in all, Boosted Perturbation Theory apparently solves the problem
of the discrepancies in between perturbative and non-perturbative
results for $Z_V$ and $Z_A$. This sounds good, also in view of the
fact that different boosted couplings basically point to consistent
results. Discrepancies are still there for $Z_S$ and $Z_P$. While
there is of course the possibility that even higher order terms should
be included, there is another explanation that could hold true. Given
the interplay of IR and UV effects, there is a possibility that
non-perturbative computations could suffer from finite volume
effects. These effects are not expected to be the same that we get
(and correct for) in our NSPT setting, but could be possibly assessed: 
more on this in the following Section. 

As a final comment, we go back to the Iwasaki case, for which
basically there was no compelling reason to go for boosted couplings
(of course one could nevertheless do it). We stress that the latter
observation could be done only in view of the control of the series at
three-loop level.

\section{Some general remarks on lattice RI'-MOM}
\label{sec:whatNEW}

There is something interesting that one can learn from our
computations, not only with respect to a comparison of perturbative
and non-perturbative results. 

First of all, we put forward a method to assess (the possible 
presence of) finite size effects. One can see that there is in
principle no reason why one should not attempt the same in the
non-perturbative case. We are actually working on this 
\cite{FinSizeAGAIN}.

Moreover, the high-loop computations which are enabled by NSPT can
provide a new handle to correct non-perturbative computations with
respect to irrelevant contributions. We have briefly sketched this in 
\cite{prLAT2013}. Quite interestingly, another group is working on the
same ideas \cite{AndreJakob}. Basically this amounts to the following 
simple recipe:

\begin{itemize}
\item One needs both a high-order NSPT computation and a
  standard non-perturbative computation.
\item First of all, one should try to assess the possible presence of 
finite size effects in both cases. We stress once again that these do
not need at all to be the same. Once assessed, they should be
corrected in both computations.
\item Once both results are corrected for (possible) finite size
  effects, one can take the irrelevant effects as estimated via the
  fitting procedure we described (here and) in \cite{NSPT_TLS} and 
subtract them from the non-perturbative data.
\end{itemize}

Subtracting irrelevant effects is by now a common practice. 
There are many approaches to this, requiring different 
combinations of perturbative computations and fitting of terms 
compliant to the lattice symmetries: see \cite{HarisGerrit} for a
recent contribution. 
In the end, our proposal is basically yet another 
variant, whose merits are worth investigating. 

\section{Conclusions and prospects}

This work is a little landmark at the end of a path that we took a few
years ago. The point we wanted to make is that the three-loop computation
of Renormalization Constants for Lattice QCD is a realistic
goal. There is in principle no sharp constraint on computing finite
constants, while for logarithmically divergent ones there is a limit 
because continuum computations are available 
at three-loop order in the RI'-MOM scheme. These results make it possible 
to derive the leading logarithmic contributions one has to account
for in the lattice regularization of the same RI'-MOM scheme. 
As for the finite parts,
Numerical Stochastic Perturbation Theory can do the job. 
All this is under control because we can assess
both finite lattice spacing (UV) and finite volume (IR) effects. 

As a general conclusion, it is fair to say that the NSPT approach to
the computation of Renormalization Constants for Lattice QCD can
provide at least two valuable contributions. First of all, it is a
completely independent approach with respect to non-perturbative
computations, with different systematic effects. From another point of
view, NSPT techniques provide a new method to correct 
non-perturbative computations with respect to irrelevant
contributions. 

Last but not least, the method we suggested for the correction of
finite size effects could be useful in non-perturbative cases as
well. 

\section*{Acknowledgments}
\par\noindent
We warmly thank Luigi Scorzato and Christian Torrero, who took part in the long-lasting
project of three-loop computation of LQCD renormalization constants
and from whom we could always have support and useful inputs. 
We are very grateful to M. Bonini, V. Lubicz, C. Tarantino, R. Frezzotti, P. Dimopoulos 
and H. Panagopoulos for all the stimulating discussions we had over
the years. M.H. warmly thanks Roberto Alfieri for having introduced him to
grid-computing. \\
This research is supported by the Research Executive Agency (REA) of the European Union under
Grant Agreement No. PITN-GA-2009-238353 (ITN STRONGnet).
We acknowledge partial support from both Italian MURST 
under contract PRIN2009 (20093BMNPR 004) and from I.N.F.N. under {\sl
  i.s. MI11} (now {\em QCDLAT}). 
We are grateful to the Research Center for Nuclear Physics and to the 
Cybermedia Center in Osaka University for the time that was 
made available to us on their computing facilities. 
We also acknowledge computer time on 
the Tramontana I.N.F.N. facility and on
the Aurora system. We thank the AuroraScience Collaboration for the latter.


\end{document}